\documentclass[11pt,letterpaper]{article}

\makeatletter
\let\blx@rerun@biber\relax
\makeatother

\usepackage[top=1in,right=1in,bottom=1in,left=1in]{geometry}

\usepackage[utf8]{inputenc}

\usepackage{csquotes}
\usepackage[english]{babel}
\usepackage[backend=biber,defernumbers=false,sortcites=true,sorting=none,maxnames=2,maxbibnames=9,maxcitenames=6]{biblatex}
\AtBeginBibliography{\small}

\usepackage{xcolor}
\definecolor{linkcolor}{HTML}{003399}

\usepackage{url}

\usepackage{graphicx}
\usepackage{amsmath}
\usepackage{amssymb}
\usepackage[T1]{fontenc}

\usepackage[font=footnotesize,labelfont=bf]{caption}

\usepackage[hypertexnames=false]{hyperref}
\hypersetup{
    colorlinks=true,
    breaklinks=true,
    citecolor=linkcolor,
    linkcolor=linkcolor,
    filecolor=magenta,
    urlcolor=linkcolor,
}

\usepackage[most]{tcolorbox}

\usepackage[nameinlink]{cleveref}

\bibliography{references}

\begin{document}

% Page header
%\markboth{Correia, Wood, Bollen \& Rocha}{Mining Social Media data}

% Title
\title{Mining social media data for biomedical signals and health-related behavior}

%Authors, affiliations address.
\author{
    \textbf{Rion Brattig Correia$^{1,2,3}$,
    Ian B. Wood$^{2}$,
    Johan Bollen$^{2}$, and
    Luis M. Rocha$^{2,1,*}$} \\
    \small $^1$Instituto Gulbenkian de Ciência, Oeiras, Portugal \\
    \small $^2$Center for Social and Biomedical Complexity, Luddy School of Informatics, Computing \& Engineering,\\
    \small Indiana University, Bloomington IN, USA \\
    \small $^3$CAPES Foundation, Ministry of Education of Brazil, Brasília DF, Brazil \\
    \small $^*$rocha@indiana.edu
}
\date{}
\maketitle

%Abstract
\begin{abstract}
\noindent Social media data has been increasingly used to study biomedical and health-related phenomena. 
From cohort level discussions of a condition to planetary level analyses of sentiment, social media has provided scientists with unprecedented amounts of data to study human behavior and response associated with a variety of health conditions and medical treatments.
Here we review recent work in mining social media for biomedical, epidemiological, and social phenomena information relevant to the multilevel complexity of human health.
We pay particular attention to topics where social media data analysis has shown the most progress, including pharmacovigilance, sentiment analysis especially for mental health, and other areas.
We also discuss a variety of innovative uses of social media data for health-related applications and important limitations in social media data access and use.
\end{abstract}

%%Keywords, etc.
{\noindent \textbf{keywords}: social media, healthcare, pharmacovigilance, sentiment analysis, biomedicine }
%%Table of Contents
%\tableofcontents

%%
%% Introduction
%%
\section{Introduction}

Humanity has crossed an important threshold in its ability to construct quantitative, large-scale characterizations of the networks of information exchanges and social interactions in human societies.
Due to the widespread digitization of behavioral and medical data, the advent of social media, and the Web’s infrastructure of large-scale knowledge storage and distribution there has been a breakthrough in our ability to characterize human social interactions, behavioral patterns, and cognitive processes, and their relationships with biomedicine and healthcare.
For instance, electronic health records of entire cities can yield valuable insights on gender and age disparities in health-care \cite{Correia:2019}, and the communication patterns of \textit{Twitter} and \textit{Instagram} help us detect the spread of flu pandemics \cite{Christakis:2010}, warning signals of drug interactions \cite{Correia:2016}, and depression \cite{Choudhury:2013a}. 

Data Science, together with artificial intelligence and complex networks and systems theory, has already enabled exciting developments in the social sciences, including the appearance of novel fields such as computational social science and digital epidemiology \cite{Lazer:2009, Salathe:2012}.
Using social media and online data, researchers in these interdisciplinary fields are tackling human  behavior and society in a large-scale quantitative manner not previously possible to study social protests \cite{Varol:2014, Kallus:2014}, fake news spread \cite{Ferrara:2016, Qiu:2017, Shao:2018}, and stock market prediction \cite{Bollen:2011}, for instance.

This approach also shows great promise in monitoring human health and disease given the newfound capability to measure the behavior of very large populations from individual self-reporting \cite{Kautz:2013}.
Indeed, building population-level observation tools allows us to study collective human behavior \cite{ChaM:2012, Ferrara:2014, Bakshy:2015, Pescosolido:2015, Fan:2018} and,
given the ability to obtain large amounts of real-World behavioral data, 
is expected to speed translational research in transformative ways \cite{Paul:2016}, including monitoring of individual and population health \cite{Hawn:2009, Seltzer:2015, Sullivan:2016, Paul:2016, MacLeod:2017, Hobbs:2016}.
This promise has been substantiated by many recent studies. Google searches have been shown to correlate with dengue spread in tropical zones \cite{ChanE:2011}. Even though the accuracy of using Google trends data alone for epidemic flu modeling has been problematic \cite{Goel:2010}, in combination with other health data it adds value \cite{Lazer:2014}.
Several studies have also shown that social media analysis is useful to track and predict disease outbreaks such as influenza \cite{Kautz:2013,Signorini:2011,Sadilek:2012}, cholera \cite{Chunara:2012}, Zika \cite{McGough:2017} and HIV \cite{Ireland:2015}, can play an important role in pharmacovigilance \cite{Hamed:2015, YangH:2013, Iyer:2014, Sarker:2015, Pain:2016, Sarker:2016, Correia:2016}, and measure public sentiment and other signals associated with depression  \cite{Choudhury:2013a, Nambisan:2015, Gkotsis:2017, Eichstaedt:2018, Choudhury:2013b} as well as public health issues such as human reproduction \cite{Wood:2017}, vaccination rates \cite{Salathe:2011, Salathe:2013} and mental disorder stigma \cite{Pescosolido:2015}.

Social media data provide an increasingly detailed large-scale record of the behavior of a considerable fraction (about 1/7\textsuperscript{th}) of the world’s population.
Since 2017, 330 million people monthly have been active users of \textit{Twitter}, making it one of the most populated global social networking platforms \cite{Statista:2019:active-twitter-users}. \textit{Instagram} currently has more than one billion monthly active users \cite{Statista:2019:active-instagram-users}.
It used to be the preferred social network among teens and young adults (12-24), but since 2016 \textit{Instagram} was surpassed by \textit{Snapchat} in this demographic \cite{Statista:2019:popular-social-network-teenagers}. \textit{Facebook}, however, still has the overall majority of active users, with 2.45 billion monthly \cite{Statista:2019:active-facebook-users}, and a total of 2.8 billion across any of the company's core products, \textit{Facebook, WhatsApp, Instagram,} and \textit{Messenger} \cite{Statista:2019:facebook}.

Biomedical and public health researchers now have the opportunity to directly measure human behavior on social media, a promise emphasized by the National Institutes of Health that consider this type of big data very relevant for biomedical research \cite{NIH:PA-14-155, NIH:PA-14-156}.
By ``social media'' we mean any user-generated content, including posts to sites such as \textit{Twitter} and \textit{Facebook}, but also comments on disease-specific or health-related sites, forums, or chats.
Most social media sites have been shown to be relevant for biomedical studies, including \textit{Twitter} \cite{Paul:2016},  \emph{Facebook} \cite{Bakshy:2015}, \emph{Flickr} \cite{ChaM:2012}, \emph{Instagram} \cite{Ferrara:2014, Correia:2016}, \emph{Reddit} \cite{Choudhury:2014, Park:2017, Zomick:2019}, and even \emph{Youtube} \cite{Fernandez-Luque:2009, Syed-Abdul:2013}.
Used together with other sources of data such as web search, mobility, scientifc publications, electronic health records, genome-wide studies, and many others, social media data helps us build population- and individual-level observation tools that can speed translational research in fundamentally new ways. 

Leveraging these kinds of data constitutes a novel opportunity to improve personalization in the face of multilevel human complexity in disease \cite{Dolley:2018, Pescosolido:2016}.
For instance, new patient-stratification principles and unknown disease correlations and comorbidities can now be revealed \cite{Fernandez-Luque:2015}. 
Moreover, social media allows a more direct measurement of the perspective of patients on disease, which is often different from that of physicians.  Social media can help both patients and practitioners to understand and reduce this disconnect \cite{Patel:2018} which is known to hinder treatment adherence \cite{Cooper:2015}.

In summary, analysis of social media data enables more accurate ``microscopes'' for individual human behavior and decision-making, as well as ``macroscopes'' for collective phenomena \cite{Lazer:2009, Borner:2011, Franca:2016}.
These micro- and macro-level observation tools can go beyond a descriptive understanding of biomedical phenomena in human populations, by enabling quantitative measurement and prediction of various processes as reviewed below.
The ability to study ``humans as their own model organism'' is now a more reasonable prospect than ever before.
Here we review recent work pertaining to the mining of social media for health-related information, that is, biomedical, epidemiological, or any social phenomena data of relevance to the multilevel complexity of human health \cite{Alber:2019}.
The review is structured as follows: \cref{ch:pharmacovigilance}, the use of social media for pharmacovigilance, including adverse drug reactions and drug-drug interactions;
\cref{ch:sentiment-analysis}, the use of sentiment analysis tools to characterize individual and population behavior, especially mental health;
\cref{ch:other}, the analysis of social media data for a wide variety of health-related applications;
%
%\Cref{ch:text-mining} covers text-mining techniques, including named-entity recognition (NER). 
%
\cref{ch:limitations}, limitations of the use of social media data; and \cref{ch:conclusion}, conclusion and final remarks.

%
% Social Media for PharmacoVigilance
%
\section{Pharmacovigilance}
\label{ch:pharmacovigilance}

It is estimated that every year the United States alone spends up to \$30.1B because of adverse drug reactions (ADR), with each individual reaction costing \$2,262 \cite{Sultana:2013}.
More than 30\% of ADR are caused by drug-drug interactions (DDI) that can occur when patients take two or more drugs concurrently (polypharmacy)  \cite{Iyer:2014}.
The DDI phenomenon is also a worldwide threat to public health \cite{Becker:2007, Correia:2019}, especially with increased polipharmacy in aging populations.

Most ADR and DDI surveillance is still conducted by analysis of physician reports to regulatory agencies, and mining databases of those reports, such as the FDA Adverse Event Reporting System (FAERS) \cite{FDA:FAERS}.
However, clinical reporting suffers from under reporting \cite{Alatawi:2017} which can be caused by clinicians failing to note adverse events or downgrading the severity of patients' symptoms \cite{Basch:2010}.
For example, it has been well documented that depression and pain are under-assessed by clinicians, under-reported by patients, and therefore under- or inappropriately managed, especially in specific cohorts such as athletes \cite{Rao:2016, Druckman:2018}. 
Even when clinicians are specifically trained or required to use screening tools for ADR, in practice these are done in a reactionary fashion at the time of a health care visit \cite{Alatawi:2017}.

\begin{figure}[!ht]
    %\centering
    \includegraphics[width=\textwidth]{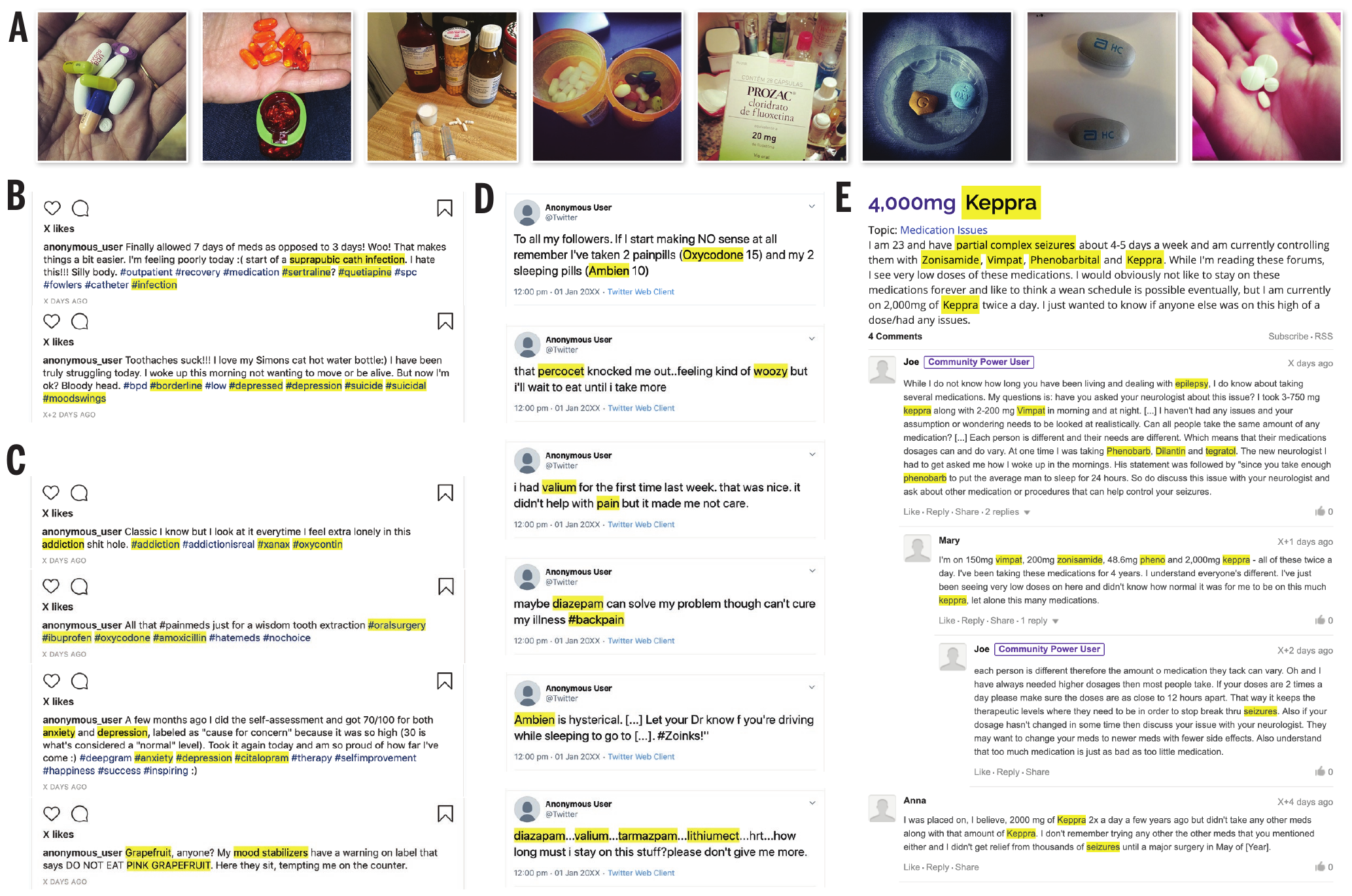}
    \caption{
        Selected sample of social media posts depicting known drug and symptom mentions.
        \textbf{A.} Photos posted to \textit{Instagram} depicting a variety of drugs.
        \textbf{B \& C.} Captions of \textit{Instagram} posts. Captions in \textbf{B} belong to the same user and were posted two days apart; the second post may contain a possible side effect from a drug administration mentioned in the first post.
        \textbf{D.} \textit{Twitter} posts containing drugs known to be abused.
        \textbf{E.} \textit{Epilepsy Foundation} forum post and comments from users asking questions and sharing experiences over drug (\textit{Keppra}) dosage.
        For all examples, usernames, number of likes, and dates were omitted for privacy; some content was modified for clarity and to maintain user anonymity; terms of pharmacovigilance interest, including drug names, natural products, and symptoms are highlighted in yellow using dictionaries developed for this problem \cite{Correia:2016,Correia:2019:thesis}.
    }
    \label{fig:pharmacovigilance}
\end{figure}

Such problems in reporting can be improved using new ways to study the ADR and DDI phenomena provided by the recent availability of large-scale online data about human behavior.
Given how many people use social media, it is likely to allow for automated proactive identification of issues as they develop rather than once they occur and are potentially severe. Thus, analysis of social media data can identify under-reported pathology associated with ADR, and further, contribute to improvements in population health (a sample of social media posts containing drug and symptoms is shown in \cref{fig:pharmacovigilance}).
For instance, it has been shown that the combination of clinical FAERS reporting and internet search logs can improve the detection accuracy of ADR by 19\% \cite{White:2014}, that discussions on \textit{Twitter} related to glucocorticoid therapy reveal that insomnia and weight gain are more common adverse events than reported in the U.K. regulator’s ADR database, and that more serious side effects are comparatively less discussed \cite{Patel:2018}.
Another study has compared patient reports of ADR on social media (various discussion forums on health-related websites) with those of clinicians on electronic health records (EHR) for the case of two the drugs aspirin and atorvastatin \cite{Topaz:2016}. 
This study found that the most frequently reported ADR in EHR matched the patients' most frequently expressed concerns on social media. However, several less frequently reported reactions in EHR were more prevalent on social media, with aspirin-induced hypoglycemia being discussed in social media only.
The observed discrepancies, and the increased accuracy and completeness in reports from social media versus those from regulator databases and EHR, has revealed that physicians and patients have different priorities \cite{Cooper:2015}. This suggests that social media provides a more complete measurement of true impact on quality of life \cite{Patel:2018} and makes it a useful complement to physician reporting.

The use of social media for pharmacovigilance is recent, but it has been receiving increasing attention in the last few years.
A review paper in 2015 has found only 24 studies almost evenly divided between manual and automated methods, and concluded that social media was likely useful for postmarketing drug surveillance \cite{Lardon:2015}.
Another review paper \cite{Sarker:2015}, also in 2015, has found 22 studies and concluded that the utility of social media data analysis for biomedicine is hindered by the difficulty of comparing methods due to the scarcity of publicly available annotated data. This has led to a shared task workshop and the ``Social Media Mining for Public Health Monitoring and Surveillance'' session at the 2016 edition of the \textit{Pacific Symposium on Biocomputing} \cite{Paul:2016}.
Shared tasks have involved the automatic classification of posts containing ADR, the extraction of related terms, and the normalization of standardized ADR lexicons \cite{Sarker:2016}.
The conference session has also attracted studies of social media data for a variety of other health-related topics: tracking emotion (see \cref{ch:sentiment-analysis}) to detect disease outbreaks \cite{Ofoghi:2016}; pharmacovigilance, including dietary supplement safety \cite{Sullivan:2016} and ADR \& DDI \cite{Correia:2016}; using machine learning to predict healthy behavior, such as diet success on publicly-shared fitness data from \textit{MyFitnessPa}l \cite{Weber:2016}, smoking cessation using \textit{Twitter} data \cite{Aphinyanaphongs:2016}, and overall well-being using \textit{Facebook} data from volunteers \cite{Sap:2016}.
Since this initial event, the shared task and workshop, currently named ``Social Media Mining for Health Applications'' (SMM4H), has been held annually, serving to bring together researchers interested in automatic methods for the collection, extraction, representation, analysis, and validation of social media data for health informatics \cite{Sarker:2017:SMM4H, Weissenbacher:2018:SMM4H, Weissenbacher:2019:SMM4H}.

Before the community was able to analyze well-known social media sites, such as \textit{Twitter} and \textit{Facebook}, most pharmacovigilance work on mining ADR from social media had been focused on social interactions in specialized health forums and message boards \cite{Chee:2009a, Leaman:2010, Nikfarjam:2011, Benton:2011, Sampathkumar:2012, Yates:2013, Patki:2014}.
One of the first to pursue this angle was Schatz' group, that has used network visualization, natural language processing, and sentiment analysis (see \cref{ch:sentiment-analysis}) to provide a qualitative ADR analysis of user comments on \textit{Yahoo Health Groups}, and has shown that it was possible to visualize and monitor drugs and their ADR in postmarketing \cite{Chee:2009a}, as well as track patient sentiment regarding particular drugs over time \cite{Chee:2009b}.

Around the same time, Gonzales' group created an ADR-focused lexicon and manually annotated a corpus of comments in support of a rule-based, lexical-matching system developed to analyze user comments in \textit{DailyStrength} (a health-focused site where users discuss personal experiences with drugs), and demonstrated that comments contain useful drug safety information \cite{Leaman:2010}.
Later the group used association rule mining to automatically extract ADR from user comments in the same platform \cite{Nikfarjam:2011}, as well as other supervised classifiers to predict if individual comments contained ADR, and a probabilistic model to infer if the \textit{DailyStrength} footprint of such predicted ADR for a given drug was likely to indicate a public health red flag \cite{Patki:2014}.

Subsequently, Benton et al. \cite{Benton:2011} used co-occurence statistics of drug-adverse effect pairs present in breast cancer message boards and compared them to drug labels of four different drugs. They found that 75-80\% of these ADR were documented on drug labels, while the rest were previously unidentified ADR for the same drugs.
Casting the extraction of (unreported) drug-event pairs in ADR as a sequence labeling problem, Sampathkumar et al. \cite{Sampathkumar:2012} used a Hidden Markov Model on patient feedback data from \url{Medications.com}, that was automatically annotated using dictionaries of drug names, side-effects, and interaction terms.

The development of several text mining and machine learning pipelines, as well as annotated corpora and lexica, for extraction and prediction of ADR from various health forums and message boards quickly ensued.
C. Yang et al.\cite{YangC:2012} used association mining and proportional reporting ratios to show that ADR can be extracted from \textit{MedHelp} user comments; this study was conducted for a small set of five known ADR (via FDA alerts) involving 10 drugs.
For the same platform, M. Yang et al. \cite{YangM:2015}, used semi-supervised text classification to filter comments likely to contain ADR, in support of an early warning pharmacovigilance system that they tested successfully, albeit with only three drugs associated with more than 500 discussion threads.
Yates et al. \cite{Yates:2013} retrieved ADR associated with breast cancer drugs by mining user reviews  from \url{askapatient.com}, \url{drugs.com}, and \url{drugratingz.com}, and produced the \textit{ADRTrace} medical term synonym set in the process.

Extraction and prediction of ADR from social media is challenging, especially because of inconsistency in the language  used to report ADR by different social groups, settings, and medical conditions \cite{Sarker:2017:SMM4H}.
Indeed, various types of evidence exist in scientific publications (e.g. in vitro, in vivo, clinical) and social media (e.g. short sentences on \textit{Twitter} and long comments on \textit{Instagram}) to report ADR and DDI.
To deal with this problem, data scientists in the field use both manual and automatic methods. The former entails manual curation by experts for each context, leading to the development of context-specific lexica and corpora, such as scientific literature reporting pharmacokinetics \cite{WuHY:2013, Kolchinsky:2015} or pharmacogenetics \cite{ZhangP:2018, WuHY:2019} studies, and tweets with medication in-take \cite{Klein:2017} or Instagram user timelines annotated with standardized drug names and symptoms \cite{Correia:2016}.  
There is also a corpus for comparative pharmacovigilance, comprised of 1000 tweets and 1000 PubMed sentences, with entities such as drugs, diseases, and symptoms \cite{Alvaro:2017}. 
Such corpora are very useful to train automatic methods to identify pharmacological relevance in both social media and scientific literature.

Automatic methods to deal with language inconsistency include automatic topic modeling and word embedding techniques that cluster similar terms according to their co-occurrence patterns with other terms \cite{Blei:2009}, typically implemented with spectral methods such as the Singular Value Decomposition (SVD) \cite{Wall:2003}.
More recently, word embeddings using neural networks, such as \textit{word2vec} \cite{Goldberg:2014, WangL:2016} have shown much promise in obtaining high-quality word similarity spaces for biomedical text \cite{JiangZ:2015} and drug lexicons for social media analysis \cite{Lavertu:2019}. Interestingly, SVD provides a linear approximation of, and insight into, what neural networks do in each layer \cite{Bermeitinger:2019} and a fast method to train them \cite{CaiC:2014}.

An example of automatic methods to deal with language inconsistency in pharmacovigilance for social media is \textit{ADRMine} \cite{Nikfarjam:2015}. It uses conditional random fields---a supervised sequence labeling classifier---to extract ADR mentions from tweets. The performance of the system is greatly enhanced by preprocessing posts (from \textit{Twitter} and \textit{DailyStrenght}) for term similarity features using unsupervised word embeddings obtained via deep learning. 
Similarly, \textit{word2vec} embedding was shown to increase the performance of automatic estimation of ADR rates of ten popular psychiatric drugs from \textit{Twitter}, \textit{Redd}it, and \textit{LiveJournal} data \cite{Nguyen:2017}, in comparison to the rates of ADR for these drugs in the SIDER database \cite{SIDER:2016}. Interestingly, the lexicon derived by \textit{word2vec} was found to leverage variants of ADR terms to deal with language inconsistency.
A drawback of using deep learning methods, however, is the need for large training corpora that limits the applicability to very commonly prescribed and discussed drugs and ADR.

%
% DDI on Social Media
%
Most work using social media data for pharmcovigilance has focused on detecting signals for single drugs and their ADR, though a few groups have studied the DDI phenomenon as well.
It started with Yang \& Yang who analyzed data from the patient discussion boards \textit{MedHelp}, \textit{PatientsLikeMe}, and \textit{DailyStrength} and focused on 13 drugs and 3 DDI pairs \cite{YangH:2013}. The study used association mining and \textit{DrugBank} as a validation database (gold standard) with good results.
From these data sets, the same group later built heterogeneous networks where nodes represented such entities as ``users,'' ``drugs,'' or ``ADR'', while edges signified ``cause'' or ``treatment.'' They went on to show that network motifs were effective in predicting DDI for an expanded set of 23 drugs, using logistic regression as a link prediction classifier \cite{YangH:2014, YangH:2015}.

Soon after, Correia, Li \& Rocha \cite{Correia:2016} were the first to study the DDI phenomenon from all the available posts on the popular social media site,  \textit{Instagram}. The group focused on seven drugs known to treat depression, collected a large data set of more than 5 million posts, and a population of almost 7 thousand users. Their study demonstrated the ability to analyze large cohorts of interest on popular social media sites. They also used a (heterogeneous) network science approach to produce a network of more than 600 drug, symptom, and natural product entities to monitor---via a web tool \cite{SyMPToM}---individual and groups of patients, ADR, DDI, and conditions of interest. The top predicted links were validated against \textit{Drugbank}, and they showed that the network approach allows for the identification and characterization of specific sub-cohorts (e.g. psoriasis patients and eating disorder groups) of relevance in the study of depression. Later on, the group expanded their work to include other epilepsy and opioid drugs, as well as analysis of \textit{Twitter} data \cite{Correia:2019:thesis}.

%
% Abuse (Opioids)
%
Recently, due to the opioid epidemic afflicting the U.S., there has been an increased interest in using social media data to understand drug abuse \cite{Kim:2017}.
Several studies have analyzed licit \cite{West:2012} (chiefly alcohol), illicit \cite{Yakushev:2014, Daniulaityte:2015b} (e.g. cocaine and marijuana), and controlled substances \cite{Hanson:2013b, Shutler:2015, Sarker:2016} (e.g. Adderall and opioids) in diverse social media sites.
Results are encouraging. For instance, analysis of \textit{Twitter} data showed that geographical activity of posts mentioning prescription opioid misuse strongly correlates with official government estimates \cite{Chary:2017}, and deep learning methods can be used to predict opiate relapse using \textit{Reddit} data \cite{YangZ:2018}.
On the other hand, an older study that considered both questionnaires and \textit{Facebook} data on five behavioral categories---including smoking, drinking, and illicit drug use---reported no significant correlation between respondent's \textit{Facebook} profiles and illicit drug use \cite{vanHoof:2014}.
Analysis of health forums on the web for this data has also shown promise. Since these forums are often anonymous, open discussion about drug abuse may be more forthcoming. One study about the drug buprenorphine---a semi-synthetic opioid effective in the treatment of opioid dependence---uncovered qualitative observations of public health interest such as increased discussion over time, perspectives on its utility, and reports of concomitant use with illicit drugs that poses a significant health risk \cite{Daniulaityte:2015a}.

Social media data could also be useful in the study of the use, potential interactions, and effects of natural products (NP) and alternative medicines, including cannabis. 
Sales of NP have nearly tripled over the past 20 years since passage of the Dietary Supplement Health and Education Act \cite{Brantley:2014}. Based on the general perception that ``natural'' means ``safe,'' the lay public often turns to NP without discussing them with their health care practitioners \cite{Blendon:2001}. Consequently, patients frequently take NP in conjunction with conventional medications, potentially triggering natural-product-drug interactions. 
The pharmacology of such products constitutes an array of ADR and DDI that are very poorly explored by biomedical research so far \cite{Brantley:2014}. This is thus an arena where social media mining could provide important novel discoveries, early warnings, and insights, particularly due to the possibility of studying large cohorts of people longitudinally \cite{Wood:2017}. 
A Preliminary study with NP and drug name dictionaries showed that it is possible to study their concomitant use longitudinally on \textit{Instagram}, and characterize associated symptoms with heterogeneous knowledge networks \cite{Correia:2016, Correia:2019:thesis}.

%%
%% Sentiment Analysis
%%
\section{Characterizing individual and collective psychological well-being}
\label{ch:sentiment-analysis}

\begin{figure}[!ht]
    %\centering
    \includegraphics[width=\textwidth]{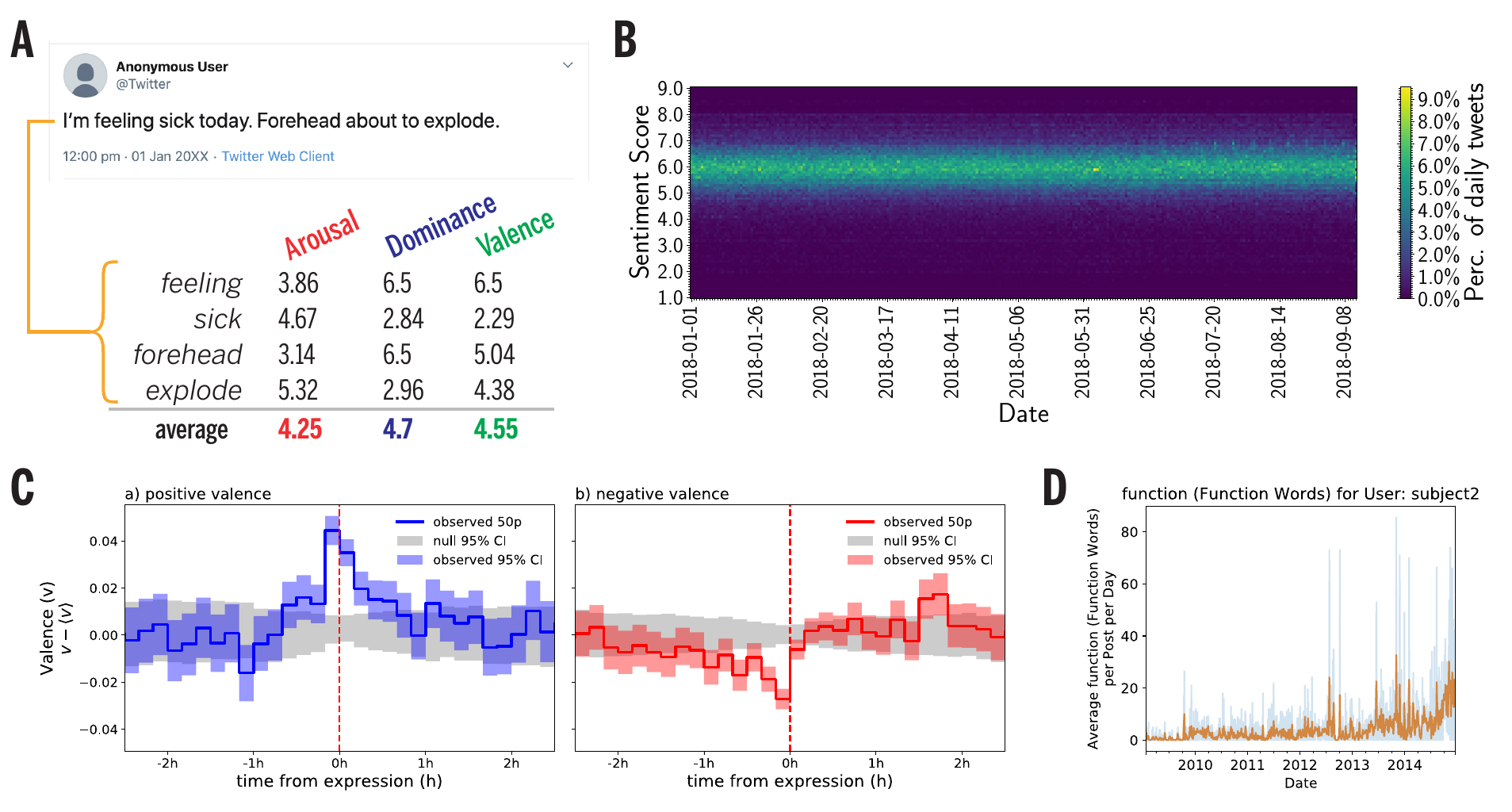}
    \caption{
        \textbf{A.} An example tweet with  its average ANEW \cite{Bradley:1999} scores for Arousal, Dominance and Valence dimensions. Only words found in the ANEW dictionary were matched to their score.
        \textbf{B}. A ``mood histogram'' timeseries, showing the per day distribution of ANEW Valence scores for a cohort of \textit{Twitter} users who self-reported as being diagnosed with depression \cite{tenThij:2019}.
        \textbf{C.} Mean-centered time series of ANEW \cite{Bradley:1999} Valence scores for a cohort of depressed users. Scores shown for 1 min increments, smoothed by a 10 min rolling average, used to study the effects of affect labeling (occurring at time 0 hour, red dashed line) on \textit{Twitter} \cite{Fan:2018}.
        \textbf{D.} Average LIWC \cite{Pennebaker:2015} functional word count for a \textit{Facebook} user timeline of a cohort of patients who died of sudden death in epilepsy (SUDEP). This young patient, like several others in cohort, showed an increase in functional words before SUDEP \cite{Wood:2019}.
    }
    \label{fig:sentimentanalysis}
\end{figure}

%% Paragraph adapted from STC grant
The psychological and social well-being of individuals and populations is an important but complex phenomenon profoundly involved in shaping health-related phenomena.
Scalable methodologies to gauge the changing mood of large populations from social media data---using natural language processing, sentiment analysis, machine learning, spectral methods, etc.---can help identify early warning indicators of lowered emotional resilience and potential health tipping points (e.g.~in the onset of mental disorders) for both epidemiological and precision health studies.

The brain is a complex system whose dynamics and function are shaped by the interactions of a large number of components. Such systems can undergo Critical Transitions (CT)\cite{Scheffer:2012}, i.e. rapid, unexpected, and difficult to reverse changes from one stable state to another. CT provide a powerful framework with which to understand and model mental health and its relation to the use of pharmaceuticals and other substances.
For instance, recent clinical longitudinal studies have provided compelling evidence that psychological CT, or tipping points, do occur in human psychological mood states. In particular, they are observed in the development of clinical depression, and are preceded by measures of critical slowing down, such as increased autocorrelation and variance, and other common antecedents, that yield useful early-warning indicators of pending CT \cite{Scheffer:2012, vandeLeemput:2014}.

Social media data is a natural source of high-resolution, large-scale, longitudinal, introspective, and behavioral data to study, monitor, and even potentially intervene before CTs occur leading to high hysteresis in efforts to return to previous equilibria.
For each individual social media user we can infer a number of important social and ethnographic factors from their online parameters, e.g. their location, language, and sex, and risk information from their statements with respect to health-risk behavior, addiction, and friendship and ``follow'' ties. 
In particular, tracking the evolving psychological mood state of social media users, over extended periods of time, along a number of relevant psychological dimensions, is now possible and widely applied \cite{Wood:2017,PakA:2010, Agarwal:2011, Zimbra:2018}. 
Indeed, social media indicators have been shown to predict the onset of depression \cite{Choudhury:2013a, Choudhury:2013b, Reece:2017, Eichstaedt:2018}. Putting one's own feelings into words on \textit{Twitter}---also known as affect labeling---can sharply reverse negative emotions, demonstrating the attenuating effects of online affect labeling on emotional dynamics and its possible use as a mood regulation strategy \cite{Fan:2018} (see \cref{fig:sentimentanalysis}).
Measuring individual and collective sentiment from social media enables the design of actionable intervention strategies to alert individuals and communities to prevent the onset of mental health issues and health risk behavior (e.g. sexual activity \cite{Wood:2017}), especially in under-served or stigmatized populations \cite{Pescosolido:2015} (see section \ref{ch:other}).

% From Johan's Proposal
The term ``sentiment analysis'' refers to a set of computational techniques that are used to measure the opinions, sentiments, evaluations, appraisals, attitudes, and emotions that people express in natural language. This sentiment can be about entities such as products, services, organizations, individuals, issues, events, topics, and their attributes, but may also include self-referential elements \cite{LiuB:2012, Pennebaker:2015}. It is also broadly defined to mean the computational treatment of opinion, mood, and subjectivity in text \cite{PangB:2008}.
The earliest studies of online sentiment relied on explicit user-defined features such as labels, ratings, reviews, etc. that were recorded as meta-data to the text \cite{PangB:2008, LiuB:2012}.
However, those features are not available for most online text, including social media posts where health-related indicators need to be inferred from unstructured, unlabeled text.
Indeed, the fundamental assumption of this approach is that individual- and population-level emotions are observable from unstructured written text.

Different methodological approaches have therefore been developed to extract sentiment indicators from text.
Some methods use natural language processing (NLP) and rely on the detection of word constructs (n-grams) in text to extract sentiment indicators with respect to an entity \cite{Nasukawa:2003}.
Other techniques classify text into positive or negative mood classes using machine learning based on annotated training sets, using e.g. support vector machines (SVM) \cite{Gamon:2004, PangB:2008, Prabowo:2009}, or naive Bayes classifiers  \cite{PakA:2010}.
Frequently, however, very good results are obtained with  lexicon-matching \cite{LiuY:2007, Dodds:2010, Esuli:2006, PangB:2002, PangB:2008}, a method that uses word lists (lexicon or dictionary) of terms that are pre-annotated with sentiment values assigned by human subjects. Lexicons of sentiment-annotated terms are obtained via a variety of methods such as expert curation and consensus, population surveys, and automatic feature construction and selection in classification tasks \cite{PangB:2008, LiuB:2012, Bradley:1999, Dodds:2015, Dodds:2010, Dodds:2011, Golder:2011, Bollen:2011:ICWSM}. This approach is particularly useful when reliability over individual text segments is less important than scalability and reliability over large-scale data sets, as is the case for social media data.
%

%% Available Sentiment Instruments
Many sentiment lexicons focus on a single dimension of measured affect, ranging from negative to positive valence (i.e. happiness).
A non-comprehensive list includes
    the \textit{General Inquirer} \cite{Stone:1962},
    the \textit{Affective Norms for English Words} (ANEW) \cite{Bradley:1999, Dodds:2010} along with several extensions and translations \cite{Redondo:2007, Soares:2012, Warriner:2013},
    \textit{Google Profile of Mood States} (GPOMS) \cite{Bollen:2011:ICWSM, Bollen:2011},
    LabMT \cite{Dodds:2011},
    SentiWordNet \cite{Esuli:2006, Baccianella:2010},
    the \textit{Lingustic Inquiry and Word Count} (LIWC) \cite{Pennebaker:2015},
    VADER \cite{Hutto:2014}, and
    \textit{OpinionFinder} \cite{Wilson:2005}.
See Box 1 for more details about these tools, especially for the categories or dimensions of sentiment that each method aims to capture. Many more sentiment analysis tools exist with extensive reviews provided by \cite{Ribeiro:2016, Reagan:2017}.

%eigenmood
These text analysis approaches, combined with large-scale social media data, have enabled the study of temporal patterns in the mood of populations at societal and planetary levels \cite{Hannak:2012, ChenX:2015, Wood:2017}. This includes studies of the changing features of language over time and geography \cite{Golder:2011, Dodds:2010}.
%In general, the application of sentiment analysis tools involves determining the central tendency of mood changes over time.

%% To say something about eigenmoods citing Yan's paper
Because collective mood estimations are derived from a collection of tweets from large and diverse populations, the resulting distributions of sentiment values can contain distinct and informative components. However, many analyses of collective mood rely on the determination of average or median sentiment over time which obscures this important information.
Spectral methods such as the Singular Value Decomposition \cite{Wall:2003} have been shown to be effective in separating sentiment components from language models in distributions of sentiment values. This approach generally: 1) removes the base sentiment contribution attributable to regular language use; and 2) extracts sentiment components (``eigenmoods'') associated with specific phenomena of interest, e.g. moods correlated with increased interest in sex \cite{Wood:2017} or depression \cite{tenThij:2019}. 
These so-called eigenmoods are components that explain a significant proportion of the variation of sentiment in time-series data instead of the average of a distribution of sentiment values reflecting prevailing language use. As such they allow more fine-grained assessments of individual and population emotions associated with health behavior of interest \cite{tenThij:2019}.

The different emotional dimensions reflected by each sentiment analysis tool have been used for specific problems relevant to health and well-being. The authors of LIWC demonstrated how its various indicators are useful to study the relation between language and a wide array of psychological problems. They were shown to reveal underlying psychological states, including the increased use of first-person singular pronouns when describing pain or trauma, verb tenses describing the immediacy of an experience, first-person plural pronouns to denote higher social status, and prepositions and conjunctions as a proxy for thought complexity, among other examples, all of which enable the measurement of individual differences \cite{Tausczik:2010}. 
For instance, textual features from the speech of student self-introductions measured by LIWC, followed by a Principal Component Analysis, were shown to be good predictors of overall academic performance. For example, the use of commas, quotes, and negative affect were positively correlated with final performance, while use of the present tense, first-person singular, home, eating, and drinking categories were negatively correlated \cite{Robinson:2013}.
LIWC has also been found useful in classifying positive vs negative affect of dream reports \cite{Nadeau:2006}, completed vs.~non-completed suicide attempts from suicide notes \cite{Pestian:2012}, measuring mood shifts and trends in large populations from social media data---e.g. feelings of sadness, anxiety, and anger---during extreme events like the September 11th World Trade Center attacks \cite{Back:2010, Back:2011} or hurricanes \cite{Kryvasheyeu:2014}.
Diurnal and seasonal rhythms in \textit{Twitter} data were found to be correlated with positive and negative sentiment in LWIC and show increased positive to negative sentiment in the morning which decreases through the day, and increased positive sentiment in days with more hours of daylight \cite{Golder:2011}.
Similarly, an analysis of more than 800 million \textit{Twitter} posts for circadian mood variations further decomposed negative mood into anger, sadness, and fatigue, finding that the latter follows an inverse pattern to the known circadian variation of concentrations of plasma cortisol---a hormone known to affect mood \cite{Dzogang:2017}.

Many other sentiment analysis tools make use of lexicons.
The ANEW (Affective Norms for English Words) lexicon \cite{Bradley:1999, Warriner:2013} consists of thousands of English words which have been rated by human subjects on three dimensions: valence, arousal, and dominance. This allow the analysis of text sentiment along distinct emotional dimensions. The ANEW was used to show that the happiness of blogs steadily increased from 2005 to 2009, exhibiting a striking rise and fall with blogger age and distance from the Earth’s equator \cite{Dodds:2010}, and that on \textit{Twitter} happiness follows a cycle that peaks on the weekend and bottoms on mid-week. \cite{Dodds:2010}.
The equally extensive LabMT lexicon was used to demonstrate that sentiment measurements are robust to tuning parameters and the removal of neutral terms \cite{Dodds:2011}.
Indeed, a comparison of different sentiment analysis tools and their performance on a number of corpora, including LabMT, ANEW, and LIWC, found that these lexical tools tend to agree on positive and negative terms, but with notable differences in performance \cite{Ribeiro:2016}. Some lexicons are dedicated to specific application areas, e.g. subjective states \cite{Wilson:2005}, whereas others are geared toward general applicability.  In general, lexical tools were found to perform well only if the sentiment lexicon covers a large enough portion of the word frequency in text and its terms are scored on a continuous scale \cite{Reagan:2017}.

Social media data has also been shown to be useful when sentiment analysis is applied to measure and address public health problems.
Qualitative content analysis of sentiment (not using automatic sentiment analysis tools) on web sites and discussion forums such as \textit{ratemds.com} revealed a positivity bias in doctor reviews \cite{Lopez:2012} and that positive reviews are associated with surgeons who have a high volume of procedures \cite{Segal:2012}.
Similar qualitative content analysis applied to \textit{Twitter} content found mostly positive views of marijuana \cite{Cavazos-Rehg:2015} with self-reports of personal use increasing when marijuana was legalized in two states \cite{Thompson:2015}.
Most early sentiment studies of the relevance of social media for public health studies are based on qualitative, manual analysis, but there has been increased interest in large-scale, automatic studies.
Using a custom sentiment analysis tool based on text classification (trained on annotated samples), \cite{Salathe:2011} studied dispositions toward flu vaccination on \textit{Twitter}. They found that information flows more often between users who share the same sentiments and that most communities are dominated by either positive or negative sentiments towards a novel vaccine (homophily) \cite{Salathe:2011}. Unfortunately for public health campaigns, they also found that negative sentiment towards vaccines spreads more easily than positive sentiment in social networks \cite{Salathe:2013}.

Choudhury et al. \cite{Choudhury:2013a} have shown that sentiment analysis tools like ANEW and LIWC are useful for analyzing the sentiment of tweets related to depression by building a large crowd-sourced corpus of tweets from individuals diagnosed with clinical depression (based on a standard psychometric instrument). They also introduced a social media depression index to characterize levels of depression in populations, and demonstrated that its predictions correlate with geographical, demographic, and seasonal patterns of depression reported by the Centers for Disease Control and Prevention (CDC).
In addition to increased negative affect, onset of depression is also found to be correlated with a decrease in social activity, stronger clustering of social interactions, heightened relational and medicinal concerns, and greater expression of religious involvement\cite{Choudhury:2013b}.

Sentiment analysis of social media data was shown to help differentiate people with post-traumatic stress disorder, depression, bipolar disorder, and seasonal affective disorder on \textit{Twitter}---and between control groups \cite{Coppersmith:2014}. The work also identified language and sentiment variables associated with each of the conditions. A similar result found that it is possible to identify linguistic and sentiment markers of schizophrenia on \textit{Twitter} \cite{WangY:2016}. 
Given that CT in mental disease are likely associated with mood changes in time that can be captured by statistical parameters like auto-correlation and variance \cite{vandeLeemput:2014}, the multidimensional, large-scale data that can be extracted from social media data, including sentiment, is likely to be of much use in the years to come.

Machine learning methods such as deep learning have been able to accurately classify social media posts according to the mental conditions to which they relate \cite{Gkotsis:2017}. This has raised the possibility of characterizing a range of mental health conditions. This approach is still relatively new and most findings are preliminary \cite{Guntuku:2017}, yet it has introduced a necessary discussion on the ethics of the utility of generating predictions respecting privacy concerns with regard to mental health-related information.

%%
%% Other Applications
%%
\section{Other promising applications}
\label{ch:other}

Social media data has been used to study a wide range of other health-related problems and has yielded promising outcomes, especially where it has been combined with other data sources.
In the area of disaster and crisis informatics, signals from social media when combined with physical sensor data, have been shown to be useful to forecast next day smog-related health hazards \cite{ChenJ:2017}.
Social media has also been used to mitigate community anxiety and the propagation of misinformation and rumours during and after environmental disasters when used effectively by credible sources like emergency response teams \cite{OhOnook:2010}.

In the area of epidemiology, social media data has been proven useful in predicting disease outbreaks such as influenza \cite{Kautz:2013, Signorini:2011, Sadilek:2012}, cholera \cite{Chunara:2012}, and Zika \cite{McGough:2017}.
In the 2015-2016 Latin American outbreak of Zika, McGough et al. \cite{McGough:2017} used a data set that combines Google searches and \textit{Twitter} data to produce predictions of weekly suspected cases up to three weeks in advance of the official publications.
Such predictions have often been correlated with the use of certain language, such as keywords or even emojis \cite{McCullom:2018}, or measured indirectly through the use sentiment analysis tools \cite{LiuB:2012}.
For instance, it has been shown that general \textit{Twitter} sentiment about vaccines correlates with CDC-estimates of vaccination rates by region \cite{Salathe:2011} (see section \ref{ch:sentiment-analysis}).
Another study has shown that higher rates of tweets containing future-oriented language (e.g., `will' and `gonna') are predictive of counties with lower HIV prevalence \cite{Ireland:2015},  and demonstrates that social media may provide an inexpensive real-time surveillance of disease outbreaks.

Social media research has also shown promise in efforts to combat stigma, offering a unique means by which to improve outcomes, benefiting healthcare providers and the public alike \cite{Bandstra:2008, LoAlto:2010, Luxton:2011, Betton:2015, Lenn:2015, Ladea:2016}. Anti-stigma advocates and government organizations already have well developed presences on web discussion boards and on social media for dozens of health conditions including obesity/body issues \cite{Lydecker:2016} and HIV \cite{Witzel:2016}, which helps raise awareness of health-related issues, including organ donations \cite{Pacheco:2017}. Efforts around epilepsy, for example, include \textit{TalkAboutIt.org}, a collaboration between actor Greg Grunberg and the Epilepsy Foundation \cite{Reynolds:2000, DeBoer:2002, Engel:2003, Schwaber:2012} and ``Out of the Shadows,'' a joint international project among the World Health Organization, the International League Against Epilepsy, and the International Bureau for Epilepsy. Efforts generally center around education and increasing public and professional awareness of epilepsy as a treatable brain disorder, and on raising public acceptability of epilepsy. 
While these efforts do not utilize data science per se, these social media platforms afford future data analyses of stigma in health.

Little data on efficacy and long-term success is available so far for epilepsy \cite{Fiest:2014} or mental health disorders \cite{Sartorius:2005}, though what exists has provided important insights. In their review, Patel et al. \cite{Patel:2015} have shown benefits provided by social media efforts, with 48\% of studies indicating positive results, 45\% undefined or neutral, and 7\% potentially harmful. Researchers accessing \textit{Twitter} have found 41\% of epilepsy-related tweets to be derogatory \cite{McNeil:2012}.  An analysis of the top ten epilepsy-related videos on \textit{Youtube} has revealed that ``real life'' or ``lived experience'' videos garner the most hits, comments, and empathetic scores but provide little information.  Videos with important health information, on the other hand, have received only neutral or negative empathy scores \cite{LoAlto:2010}.  As a contributing factor, concerns about privacy and others' reactions limit respondents' willingness to access and engage content on a website \cite{Payton:2016}. In one of the only network-based studies so far, Silenzio et al. \cite{Silenzio:2009} have found that mapping social network interactions reveals that some individuals on social media may be more important to efforts to spread anti-stigma interventions. Additional research in this area is clearly needed.

The study of health-related issues around human sexuality can also be improved by analysis of web search, social media discourse, and health forum data, especially those platforms that provide anonymity such as \textit{Reddit} \cite{Choudhury:2014}.
For instance, web search and \textit{Twitter} data have been instrumental in clarifying competing hypotheses about the cyclic sexual and reproductive behavior of human populations. Rather than an evolutionary adaptation to the annual solar cycle, analysis of planet-wide data has suggested that observed birth cycles are likely a cultural phenomenon---since increased interest in sex is correlated with specific emotions, characteristic of major cultural and religious celebrations \cite{Wood:2017}.

On the medical side of human reproduction issues, pregnant women have frequently turned to the Web and social media for reassurance on the normalcy of their pregnancies and to gather information on normal pregnancy symptoms, pregnancy complications, birth, and labor \cite{SongF:2012, Lupton:2016}. For first-time mothers in particular, social media platforms have appeared to be the preferred mechanisms for obtaining important information during the antepartum and postpartum periods'' \cite{Harpel:2018, Asiodu:2015}. Posting status updates and photos on social media appears to have provided pregnant women with a sense of connection with their peers, as well as with their own unborn babies \cite{Lupton:2016, Bartholomew:2012}.
Considering, in addition, the numbers of legal and illegal drug users as described above, social media platforms appear to be untapped sources of large-scale data on under-reported, population-level risk for neonatal and related conditions, such as neonatal abstinence syndrome. Social media signals may be effective resources to model the pharmacological, phenotypical, and psychosocial markers associated with drug use during pregnancy, and may lead to better early problem warnings and prevention strategies.

Other measures that are known to correlate with health outcomes have also been investigated. For instance, social media deviations in diurnal rhythms, mobility patterns, and communication styles across regions have been included in a model that produces an accurate reconstruction of regional unemployment incidence \cite{Llorente:2015}.
Also, the potential of social media to predict severe health outcomes in epilepsy is preliminary but promising. Sudden Unexpected Death in Epilepsy (SUDEP), for example, remains a leading cause of death in people with  epilepsy. A small study of the \textit{Facebook} timelines of people who died in this way was conducted to identify potential behavioral signs preceding a SUDEP, and has suggested that prior to dying a majority of the subjects wrote more text than they had previously on social media \cite{Wood:2019}.

%%
%% Limitations
%%
\section{Limitations}
\label{ch:limitations}

Social media can yield useful healthcare information, but there are inherent limitations to its use for biomedical applications. On the positive side, because analysis takes place after the data is recorded, social media analytics in general avoids experimenter and social conformity bias. Social media data is a type of real-World data \cite{Ramagopalan:2020} that allows for very large-scale population samples which surpass those of traditional social science and clinical trial approaches by several orders of magnitude. In fact, \textit{Twitter} offers strong opportunities for academic research given its public nature, real-time communication, and user population that approaches significant pluralities of the world's population. This is also the case for other social media platforms such as \textit{Facebook}, \textit{Instagram}, and \textit{Reddit}. On the other hand, the data collected is frequently lacking demographic indicators and ground-truth, possibly resulting in biased or poorly representative samples---particularly when contrasted with precisely defined inclusion and exclusion criteria of randomised controlled trials (RCT) \cite{Ramagopalan:2020}. 
In this section we provide a short overview of the literature related to the challenges of deriving valid and reliable indicators of human behavior from social media data and how they can be mitigated.

In spite of scale, social media data generally entail self-selected samples, since subjects are free to choose when to participate and what content to submit. This bias is compounded by a mix of access restrictions imposed by social media platforms \cite{Pfeffer:2018}. Researchers are, as a result, prone to use so-called ``convenience samples'', i.e.~social media data sets that are, due to standardization efforts, more widespread, accessible, and convenient to use, though they may not represent the wider population. Combined, these biases may lead to samples that do not validly or completely represent human behavior and diversity.

The content of social media data may also be subject to lexical bias \cite{Dodds:2015} that could cause sentiment data to over-represent positive sentiment. In addition, platform-specific factors may alter user behavior \cite{Ruths:2014, Pfeffer:2018} and lead to bias in subsequent data analysis. In fact, users may be encouraged to engage in profile and reputation management by establishing different online personas to highlight their individuality and qualities that are perceived as desirable \cite{Jensen:2017}.

Privacy issues and algorithmic bias may also lead to mischaracterization of human factors. The behavior of most social media users is profoundly shaped by interface designs and, increasingly, algorithmic factors, e.g.~the use of machine learning services for recommendations of social relations and relevant content. Non-human participants such as bots are, furthermore, widespread in some social media sites, e.g. \textit{Twitter} \cite{Ferrara:2016}.
Moreover, exogenous events such as polarized elections, will trigger individual and global sentiment changes, cause discourse polarization, and bring about temporary deviations from baseline social-linkage dynamics. The social-economic-political context in which the particular social media data was recorded will therefore play an important role.

Given all these potential population biases, mining social media for healthcare information relevant to the broader human population requires a careful consideration of the multi-leveled complexity of human health \cite{Alber:2019}, in which social and behavioral contexts play a critical role \cite{Shaban-Nejad:2018}.

Perhaps one of the most important issues with social media mining is the difficulty in establishing sample validity ane precise inclusion and exclusion criteria. Primarily, two sources of bias impact harvested social media data: \emph{sampling bias} and \emph{algorithm bias}. Sampling bias means that researchers cannot treat sampled social media data, e.g. \textit{Twitter's} 1\% sample, as true, random, human population samples. This affects efforts to build valid cohorts and make generalizations from analytics \cite{Pfeffer:2018}. Samples cannot be balanced because of the lack of ground truth with respect to user demographics.
Furthermore, the demographics of social media sites can vary broadly. In a survey of social media usage among U.S. adults, 43\% of women said they have used \textit{Instagram} at least once, while for men this number was only 31\%. Similarly, Hispanics appear underrepresented on LinkedIn---only 16\% said they have used the platform as compared to 24\% of Whites and 28\% of Blacks. At the same time, Hispanics appear to be the largest demographic on WhatsApp---42\% as compared to 24\% of Whites and 13\% of Blacks \cite{PewResearch:2019:Share-of-US}.

Sampling bias can be accentuated when sub-cohorts of social media users are used to draw geographical inferences, e.g.~when particular keyterms in user content are used to infer location. Such sub-samples may vary considerably in terms of the degree to which they represent an unbiased sample.
Future research using social media data must be able to benefit from the large-scale nature of this real-World data, while specifying more precise inclusion and exclusion criteria, as used in RCT to avoid sample biases \cite{Ramagopalan:2020}. Getting there requires the ability to stratify social media user cohorts by using more fine-tuned machine learning, as well as via greater collaboration with and openness from social media providers. %
It is encouraging that \textit{Twitter} data has been shown to match census and mobile phone data in geographical grids down to one square kilometer resolution \cite{Lenormand:2014}. Indeed, machine learning methods can be used on \textit{Twitter} data to automatically track the incidence of health states in a population \cite{Prieto:2014}. Moreover, User demographics such as age and gender can be estimated from user content with reasonable accuracy \cite{Fan:2018}.

% algorithmic bias
In addition to sample bias, it is important to be aware of algorithmic biases that result from interface design, policies, and incentives associated with social media platforms. Since company revenues are tied to targeted advertisement, social media algorithms are tailored for navigation retention and profile building. These algorithms are highly dynamic, proprietary, and secret, all of which have consequences for research reproducibility \cite{Ruths:2014}. Most researchers, like users, are largely unaware of how platforms filter timelines and other user information \cite{Luckerson:2015}. Therefore, greater openness on the part of social media companies, perhaps encouraged or mandated by public policy, is needed to increase the utility of this data for biomedicine.

In addition to sample and algorithmic biases, sentiment analysis can be manipulated by third parties through the injection of tweets \cite{Pfeffer:2018}, i.e. the deliberate insertion of tweets containing specific words known to affect sentiment tools, e.g.~to boost general sentiment during a political debate. These efforts can be difficult to detect and mitigate since they affect the sample \textit{a priori}, before a researcher can apply efforts to unbias their sample and address sample validity.
Indeed the extraction of emotional and social indicators from social media is fraught with difficulty. 
Users may indirectly disclose mood states, sentiment, health behavior, and diet, but rarely do so explicitly \cite{Fan:2018}. Social media users furthermore favor an an idiomatic style and vernacular that is difficult to analyze with traditional Natural Language Processing (NLP) tools and supervised Machine Learning (ML) algorithms. Applications of the latter are hampered by the lack of vetted ``ground truth'' data sets and the highly dynamic nature of underlying emotional processes.
Additional difficulties in analyzing social media discourse include the use of sarcasm (particularly toward effects of specific drugs), subjective opinion, and the polarity of a sentiment-laden word or phrase in context \cite{PangB:2008,Hutto:2014}.
For instance, social media users may use the term ``Prozac'' in a variety of idiosyncratic usages, but not necessarily because they are actually administering the drug.

Users revealing sensitive personal information about others, as well as information pertinent to their social relationships, raises serious privacy concerns. In fact, data from eight to nine social media relations of an individual is sufficient to predict that individual's characteristics just as well as from their own information \cite{Bagrow:2019}. In other words, privacy concerns are not just a matter of what users reveal about themselves, but what their social relations (unwittingly) reveal about them. Some users are aware of this phenomenon which lowers their motivation and willingness to participate in studies from social media data \cite{Payton:2016}.

Another limitation, finally, is the danger of overfitting in subsequent analysis. Because of data availability and privacy issues, information on specific cohorts is derived from indicators that are in turn derived from the content they generate. This will favor certain content and cohorts, possibly leading to models that overfit the data and generalize poorly \cite{Ginsberg:2009,Lazer:2014}.

%%
%% Conclusion
%%
\section{Conclusion}
\label{ch:conclusion}

The studies reviewed in Pharmacovigillance (section \ref{ch:pharmacovigilance}) show that social media users discuss a wide variety of personal and medical issues, e.g. their medical conditions, prognoses, medications, treatments, and quality of life, including improvements and adverse effects they experience from pharmacological treatments \cite{Hang:2012, Hamed:2015, Correia:2016, Patel:2018}.
This collective discourse in turn can be monitored for early warnings of potential ADR, and also to identify and characterize  under-reported, population-level pathology associated with therapies and DDI that are most relevant to patients \cite{Correia:2016, Paul:2016, Hamed:2015,Sarker:2015,Dolley:2018}.
The new data-enabled modes of pharmacovigilance that social media affords are likely to be particularly relevant for patient-centered management and prevention of chronic diseases \cite{myAURA}, such as epilepsy \cite{Miller:2017}, and inflamatory diseases \cite{Patel:2018}, which continue to be the chief health care problem in the USA \cite{CDC:2019:NCCDPHP}.
The inclusion of signals from, and engagement with, social media in patient-centered personal health library services that can store, recommend and display individualized content to users is expected to significantly improve for chronic disease self-management \cite{myAURA}, which is known to significantly lower disease severity and the number of unhealthy days, and improve quality of life \cite{Grady:2014, Webel:2013}.

It is clear that disease prevention is increasingly becoming a matter of the mitigation of individual lifestyles and decision-making, which are subject to a range of cognitive, emotional and social factors that have until now been difficult to assess with sufficient accuracy and scale. 
% (...)
An understanding of the emotional and social factors that contribute to the emergence of public health issues is crucial for efficient mitigation strategies.
As we describe in section \ref{ch:sentiment-analysis}, there is already a substantial body of literature on characterizing psychological well-being, especially by measuring individual and collective sentiment and other social interactions online. These methods have been particularly effective when used in combination with other sources of health and human behavior data, from physical sensors, mobility patterns, electronic health records, and more precise physiological data.

The methodologies we cover also fall in the area of computational social science, which is presently focused on establishing the methodological framework to monitor societal phenomena from large-scale social media data---the aforementioned  social ``macroscopes.'' 
For this methodology to be relevant in the prevention of disease and improvement of public health, researchers need to move from descriptive inductive modes of analysis to explanatory models with predictions and testable hypotheses.
In particular, researchers need to establish social media not just as a tool for observation, but also as the foundation for explanatory models of the generative factors in health behavior and outcomes, of the type that computational and complexity sciences are already producing, e.g. in molecular and organismal biology \cite{Stoeger:2018, Cassidy:2019}.

There is reason to be optimistic about our ability to reach such predictive explanatory models since we know from psychological research that emotions play an significant role in human decision-making \cite{Dolan:2002, Damasio:1994, Kahneman:1979}. Behavioral finance in particular, for example, has provided evidence that financial decisions are significantly driven by emotion and mood \cite{Nofsinger:2005}. 
Hence it is reasonable to assume that online mood and sentiment, as well as all social media analysis we review, may be used to predict health behaviors, and can therefore be used to predict individual as well as societal health outcomes.

The literature we review above also points to a newfound ability to use social media data for improved well-being of small, specific cohorts and even individuals using precise characterizations and interventions. These may include pharmacological warnings, chronic disease patient-centered management, and mental disorder assistance. 
For instance, donated timelines from at-risk-of-suicide inviduals can help machine learning models recognize early-warning symptoms of depression and suicide \cite{Ruiz:2016}.

Despite its proven importance to the specific goal of improving human health, social media data has been increasingly difficult to collect. 
Only a few social media data sources remain open for scientists; many previously accessible sites are now almost entirely sealed from researchers, which is surprising given that the data is generated by and for its users, not the platforms which mostly serve a mediating function.
These limitations explain why most of the work reported has focused on \textit{Twitter}, which remains open for data analysis. Nonetheless, other social networks have been shown to be useful for biomedicine, including \textit{Facebook} \cite{Bakshy:2015}, \textit{Flickr} \cite{ChaM:2012}, \textit{Instagram} \cite{Ferrara:2014, Correia:2016}, \textit{Reddit} \cite{Choudhury:2014, Park:2017, Zomick:2019}, and even \textit{Youtube} \cite{Fernandez-Luque:2009, Syed-Abdul:2013}. 

It is possible that government policies may be leveraged to ensure accessibility to these important data sources, which could be considered a public good to be regulated much like publicly funded scientific publication data \cite{Margolis:2014, PlanS:2019}.
This would help improve the sample and algorithmic limitations discussed in section \ref{ch:limitations}, allowing this large-scale, real-World data to better identify health factors that more expensive clinical trials cannot due to their smaller scale and cost.
We hope that our review contributes to establishing the importance of social media data for biomedical research and demonstrates the need to make this data more accessible in general to scientific research.

%%
%% Acknowledgments
%%
\begingroup
    \footnotesize
    \vspace{5mm}
    \noindent
    \textbf{Acknowledgments.}
    The authors would like to thank Deborah Rocha for editing this review and Marijn ten Thij for providing plots for figures.
    RBC was funded by CAPES Foundation, grant 18668127, and Fundação para a Ciência e a Tecnologia, grant PTDC-MEC-AND-30221-2017.
    JB thanks the Economic Development Agency (EDA/ED17HDQ3120040), the National Science Foundation (SMA/SME \#1636636), the SparcS Center of Wageningen University, and the ISI Foundation for their support.
    LMR was funded by the National Institutes of Health, National Library of Medicine, grants 1R01LM012832-01 and 01LM011945-01, and by NSF-NRT grant 1735095 "Interdisciplinary Training in Complex Networks and Systems" and Fundação Luso-Americana para o Desenvolvimento and National Science Foundation, 276/2016.
    The funders had no role in study design, data collection and analysis, decision to publish, or preparation of the manuscript.
    \normalsize
\endgroup

%% Print References
\printbibliography

%%
%% Box about some important Sentiment Analysis Tools
%%
\begin{tcolorbox}[breakable,title=Box 1 - Sentiment Analysis Tools]
    
%
%% General Inquirer
The \textbf{General Inquirer} \cite{Stone:1962} was developed as a tool to organize nonnumerical data, and tag words in a text across various categories, and allow the text to be organized according to such tags. The system started as a general-purpose tool with a dictionary of categories over the 3000 most common English words and a few hundred words of interest to a behavioral scientist. The categories included “Persons”, “Behavorial Processes”, “Psychological States”, and more, for the purpose of content analysis to trace psychological themes over a series of group discussions It has since grown to include the “Harvard IV-4” and “Lasswell” content analysis dictionaries as well, for a total of 198 categories \cite{Stone:2019:tutorial}.

%
%% ANEW
\textbf{The Affective Norms for English Words} includes ratings from 1 to 9 for 1034 words along three mood dimensions: valence from unhappy to happy, arousal from calm to excited, and dominance from controlled to in-control. These ratings were collected from surveys given to undergraduates in a psychology class using a 9-point Likert-like scale \cite{Bradley:1999}.
It has been used as a basis for a number of new dictionaries, including an extension to nearly 14,000 words \cite{Warriner:2013}, a translation to Spanish \cite{Redondo:2007}, European Portuguese \cite{Soares:2012}, among others.

%
%% GPOMS
The \textbf{Google Profile of Mood States} is an extension of the Profile of Mood States (POMS), a test of self-reported Likert-scale questions measuring 6 underlying dimensions of mood: Tension or Anxiety, Depression or Dejection, Anger or Hostility, Vigor or Activity, Fatigue or Inertia, and Confusion or Bewilderment \cite{McNair:1981}.
GPOMS tries to translate the questionaire to a dictionary suitable for sentiment analysis of large-scale social media data. This tool extended the original 72 terms in the POMS questionnaire to a dictionary of 964 words by looking at co-occurrences in Google’s 4, and 5-gram corpora. These terms correspond to moods across 6 categories: calm, alert, sure, vital, kind, and happy \cite{Bollen:2011:ICWSM}.

%
%% LabMT
\textbf{LabMT} used Amazon’s Mechanical Turk to send out ANEW-like surveys ranking 1,000s of words on a 9-point scale from sad to happy, collecting at least 50 ratings for each word. Initially, LabMT was comprised of 10,222 English words found by merging the 5,000 most used words in each of four corpora: Google Books, \textit{Twitter}, music lyrics, and the New York Times\cite{Dodds:2011}. This has since been extended to include 10 languages with about 10,000 words each collected across 24 corpora \cite{Dodds:2015}.

%
%% LIWC
\textbf{Linguistic Inquiry and Word Count}, LIWC (pronounced “Luke”), is a software tool for text analysis whose first version was released publicly in 2001 and has been actively sup- ported and widely used since \cite{Tausczik:2010, Pennebaker:2015}. LIWC was developed by a number of judges through a well-documented procedure. Judges independently created lists of words, tested for consistent categorization between a majority of judges, uncommon words not present in a variety of corpora (blogs, novels, spoken language studies, etc.) were removed, internal consistency was evaluated with a corrected Cronbach’s alpha calculation, and external validity was tested through psychological studies, including writing prompts for students. The latest version of the software, LIWC2015, has dictionaries containing nearly 6,400 words and produces outputs across about 90 categories, including positive and negative emotion, and also pronouns, articles, congitive processes, time focus, personal concerns, and informal language among others \cite{Pennebaker:2015}.

%
%% SentiWordNet
\textbf{SentiWordNet} \cite{Esuli:2006, Baccianella:2010} is a dictionary assigning words values from 0 to 1 along three dimensions: Objective, Positive, and Negative, such that all values sum to one for each word. SentiWordNet was built on synsets, groups of synonymous words, from WordNet \cite{Miller:1995} and the lexical relationships between them. A committee of ternary classifiers were trained in a semi-supervised fashion. Starting from a small set of positive or negative labeled seeds, labels were propagated to related synsets within various radii, and various supervised classifiers were trained on these sets. The final values for each word/synset are determined by the proportion of classifiers labeling the synset as objective, positive, or negative, with random walk dynamics further refining values \cite{Baccianella:2010}.

%
%% VADER
\textbf{VADER} \cite{Hutto:2014} is a tool for measuring the extent of positive or negative sentiment with more than a dictionary, and is readily available as part of the natural language toolkit for python. In addition to dictionary-based sentiment scores, VADER looks at other words in a sentence modifies sentiment scores based on 5 simple rules, namely the presence of exclamations (e.g. “!!!”), capitalization, adverbs (e.g. “very”), negations (e.g. “not”), and contrastive conjunctions (e.g. “but”).

%
%% Opinion Finder
\textbf{OpinionFinder.} \cite{Wilson:2005} is a full processing pipeline, first tokenizing a document, and then using a series of classifiers trained on various corpora to find subjective statements, find speech events, identify opinion source, identify expressions of sentiment, and finally to identify the expression as positive or negative.

%
%% OTHER
In addition to the listed sentiment analysis tools above, other work suggests modifications of sentiment scores through context, for example, through compositional rules, to modify sentiment scores from sentence parse trees \cite{Moilanen:2007}. Other extensions to dictionary-based sentiment analysis involve techniques to build features for traditional machine learning classifiers (Naive Bayes, SVM, etc.) on top of or in lieu of lexical scores \cite{Hutto:2014}. Such features include word modifications, grammatical position, sentence-level, and document-level features \cite{Wilson:2005}; semantic features identifying the type of entity discussed (person, place, etc.) \cite{Saif:2012}; features from a hidden markov model latent dirichlet allocation analysis \cite{Duric:2011}; or comparing the parse trees of text from different classes through boosting methods \cite{Kudo:2004}. These methods will not be explored in detail here, for most are not available out-of-the-box, and must be trained for specific tasks.

\end{tcolorbox}

%%
%% End of BOX
%%

\end{document}